\documentclass[11pt,a4paper,onecolumn,numbers]{article}

\usepackage{arxiv_icrc2011}

\title{Mass sensitivity in the radio lateral distribution function}

\authors{
W.D.~Apel$^{2}$,
J.C.~Arteaga$^{1,14}$,
L.~B\"ahren$^{3}$,
K.~Bekk$^{2}$,
M.~Bertaina$^{4}$,
P.L.~Biermann$^{5}$,
J.~Bl\"umer$^{1,2}$,
H.~Bozdog$^{2}$,
I.M.~Brancus$^{6}$,
P.~Buchholz$^{7}$,
E.~Cantoni$^{4,8}$,
A.~Chiavassa$^{4}$,
K.~Daumiller$^{2}$,
V.~de~Souza$^{1,15}$,
F.~Di~Pierro$^{4}$,
P.~Doll$^{2}$,
R.~Engel$^{2}$,
H.~Falcke$^{3,9,5}$,
M. Finger$^{1}$, 
B.~Fuchs$^{1}$,
D.~Fuhrmann$^{10}$,
H.~Gemmeke$^{11}$,
C.~Grupen$^{7}$,
A.~Haungs$^{2}$,
D.~Heck$^{2}$,
J.R.~H\"orandel$^{3}$,
A.~Horneffer$^{5}$,
D.~Huber$^{1}$,
T.~Huege$^{2}$,
P.G.~Isar$^{2,16}$,
K.-H.~Kampert$^{10}$,
D.~Kang$^{1}$, 
O.~Kr\"omer$^{11}$,
J.~Kuijpers$^{3}$,
K.~Link$^{1}$, 
P.~{\L}uczak$^{12}$,
M.~Ludwig$^{1}$,
H.J.~Mathes$^{2}$,
M.~Melissas$^{1}$,
C.~Morello$^{8}$,
J.~Oehlschl\"ager$^{1}$,
N.~Palmieri$^{1*}$,
T.~Pierog$^{2}$,
J.~Rautenberg$^{10}$,
H.~Rebel$^{2}$,
M.~Roth$^{2}$,
C.~R\"uhle$^{11}$,
A.~Saftoiu$^{6}$,
H.~Schieler$^{2}$,
A.~Schmidt$^{11}$,
F.G.~Schr\"oder$^{2}$,
O.~Sima$^{13}$,
G.~Toma$^{6}$,
G.C.~Trinchero$^{8}$,
A.~Weindl$^{2}$,
J.~Wochele$^{2}$,
M.~Wommer$^{2}$,
J.~Zabierowski$^{12}$,
J.A.~Zensus$^{5}$
}
\afiliations{
$^1$ Institut f\"ur Experimentelle Kernphysik, KIT - Karlsruher Institut f\"ur Technologie, Germany\\
$^2$ Institut f\"ur Kernphysik, KIT - Karlsruher Institut f\"ur Technologie, Germany\\
$^3$ Radboud University Nijmegen, Department of Astrophysics, The Netherlands\\
$^4$ Dipartimento di Fisica Generale dell' Universit\`a di Torino, Italy\\
$^5$ Max-Planck-Institut f\"ur Radioastronomie Bonn, Germany\\
$^6$ National Institute of Physics and Nuclear Engineering, Bucharest, Romania\\
$^7$ Fachbereich Physik, Universit\"at Siegen, Germany\\
$^8$ Istituto di Fisica dello Spazio Interplanetario, INAF Torino, Italy\\
$^9$ ASTRON, Dwingeloo, The Netherlands\\
$^{10}$ Fachbereich Physik, Universit\"at Wuppertal, Germany\\
$^{11}$ Institut f\"ur Prozessdatenverarbeitung und Elektronik, KIT - Karlsruher Institut f\"ur Technologie, Germany\\
$^{12}$ Soltan Institute for Nuclear Studies, Lodz, Poland\\
$^{13}$ Department of Physics, University of Bucharest, Bucharest, Romania\\
\scriptsize{
$^{14}$ now at: Univ Michoacana, Morelia, Mexico;
$^{15}$ now at: Univ S$\tilde{a}$o Paulo, Inst. de F\'{\i}sica de S\~ao Carlos, Brasil; 
$^{16}$ now at: Inst. Space Sciences, Bucharest, Romania 
}
}
\email{$^{*}$Author: Nunzia Palmieri -- Nunzia.Palmieri@kit.edu}

\abstract{
Measuring the mass composition of ultra-high energy cosmic rays is one of the main tasks in the cosmic rays field. Here we are exploring the composition signature in the coherent electromagnetic emission from extensive air showers, detected in the MHz frequency range. One of the experiments that successfully detects radio events in the frequency band of 40-80 MHz is the LOPES experiment at KIT.
It is a digital interferometric antenna array and has the important advantage of taking data in coincidence with the particle detector array KASCADE-Grande. A possible method to look at the composition signature in the radio data, predicted by simulations, concerns the radio lateral distribution function, since its slope is strongly correlated with Xmax. Recent comparison between REAS3 simulations and LOPES data showed a significantly improved agreement in the lateral distribution function and for this reason an analysis on a possible LOPES mass signature through the slope method is promising. Trying to reproduce a realistic case, proton and iron showers are simulated with REAS3 using the LOPES selection information as input parameters. The obtained radio lateral distribution slope is analyzed in detail. The lateral slope method to look at the composition signature in the radio data is shown here and a possible signature of mass composition in the LOPES data is discussed.}

\begin{document}
\maketitle

\section{Introduction}
\begin{figure}[h!]
  \vspace{2mm}
  \centering
  \includegraphics[width=5.in,height=3.in]{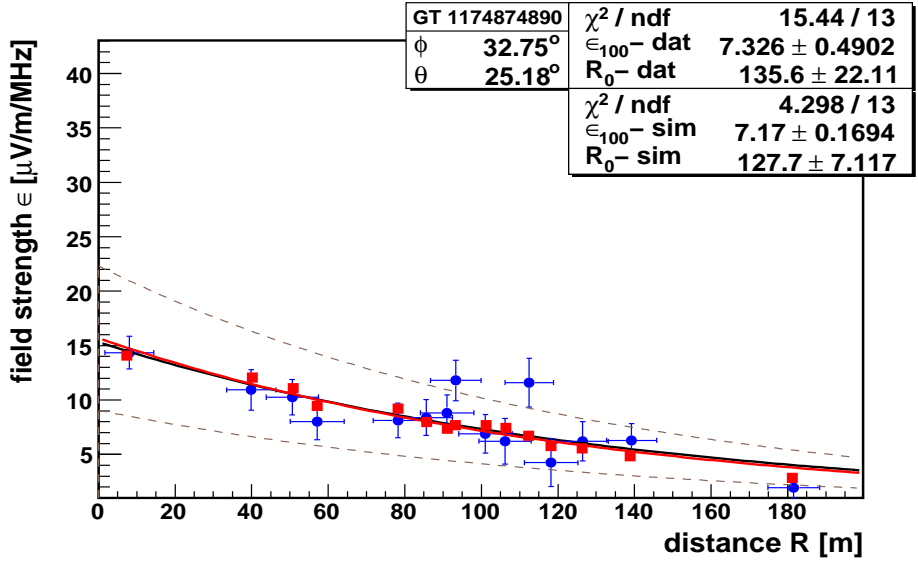}
  \caption{Lateral distribution function for one LOPES event (points) and REAS3 simulation (squares). The fitting function used is eq.\ref{expf}. The dotted lines indicates the absolute calibration and the KASCADE energy reconstruction uncertainties, not included in the simulations.}
  \label{LDFcomp}
\end{figure}

Radio detection of the cosmic ray air showers made considerable progress in the last few years. Nevertheless its sensitivity to the primary composition has not yet been tested experimentally. 
Different simulations try to model the radio emission from cosmic ray air showers, and they mostly agree on the geomagnetic induction effect as the main emission mechanism \cite{REAS3,MGMR}. A minor contribution comes from the charge-excess radiation, which influences the radio lateral distribution function (LDF) \cite{REAS3,MGMRc}. 
The LDF slope is predicted to be an indicator of the primary mass \cite{REAS2c}. Iron nuclei interact earlier in the atmosphere, so their X$_{\mathrm{max}}$, i.e. the atmospheric depth of the shower maximum, is further away from the observer compared to the X$_{\mathrm{max}}$ of lighter cosmic rays. This causes an expectation for the radio lateral distribution function slope to be flatter for heavier and steeper for lighter primaries. 
The good agreement between LOPES data and simulations with the new REAS3 code, which includes the total complexity of the radio air shower emission, so also the charge excess contribution, suggested to revisit the radio lateral slope analysis, investigated in \cite{REAS2c} with the REAS2 code. For this, REAS3 simulations for a selection of LOPES events are analyzed in direct comparison with the measured data.

\section{LOPES simulated events}
LOPES is a digital interferometric radio antenna array, optimized for the detection of cosmic rays with a primary energy around 10$^{17}$eV \cite{lopes}. It is placed at KIT, Germany, and profits from the reconstructed air shower parameters of both the particle detector arrays KASCADE and KASCADE-Grande \cite{kascade,kg}. The radio data selected for the slope analysis have been measured with the LOPES30 and LOPESpol setups \cite{timArena}. The first consisted of 30 calibrated dipole antennas all oriented in the east-west direction, while the second used only 15 antennas aligned in the east-west direction, all operating in the frequency range of 40-80 MHz. The core position of the shower used in this selection is required to be at a distance of at most 90 m from the center of the LOPES array and is reconstructed by KASCADE, with the high resolution of 4 m \cite{kascade}. High signal-to-noise and high coherency for the radio signal in the antennas is required. Further qualitative cuts demand a good fit for the lateral distribution function, i.e. small $\chi^{2}$ and good reconstructed fitting parameters. 
235 events are so selected, with a zenith angle between 0-40 degree. In fig.\ref{LDFcomp} the LDF for one detected  LOPES event (points) and its REAS3 simulation (squares) are shown, where R is the distance to the shower axis, and $\epsilon$ the electric field strength detected or simulated in each antenna, divided by the effective bandwidth (31 MHz). The points are then fitted with an exponential function 
\begin{equation}
\epsilon (R) \simeq \epsilon_{100} \mathrm{exp}(-(R -100 \mathrm{m}) / R_{0})
   \label{expf}
\end{equation}
with two free parameters, the scale parameter R$_{0}$, which describes the lateral slope, and $\epsilon_{100}$, the electric field value at 100 m from the shower axis.\\
Thanks to the improvements made in modeling the radio emission mechanism in REAS3, a good agreement can be seen between the measured and simulated LDFs for almost all the events.\\
For the slope-method analysis, the REAS3 simulations are made to reproduce a realistic case, therefore the information about the LOPES selected events is used as input parameters. The core position, the number of muons (N$\mu$), the primary energy and the arrival direction of the air shower are all directly taken from KASCADE and KASCADE-Grande. Moreover the observer positions are simulated to correspond to the real LOPES antenna position in the field with respect to the core of the shower.\\ 
Before simulating the radio emission with the REAS3 code, the measured N$\mu$ is used for the pre-selection of the CORSIKA shower. 200 CONEX showers for proton and 100 for iron are simulated with QGSJetII and UrQMQ respectively for high and low energy interaction\cite{cors,marianne}. Among all, it is chosen the CONEX shower which can best reproduce the measured N$\mu$. In this way a specific shower similar to what has been detected by KASCADE is used.


\section{Slope-method based on REAS3 simulations}
\begin{figure}[h!]
  \vspace{5mm}
  \centering
  \includegraphics[width=5.in]{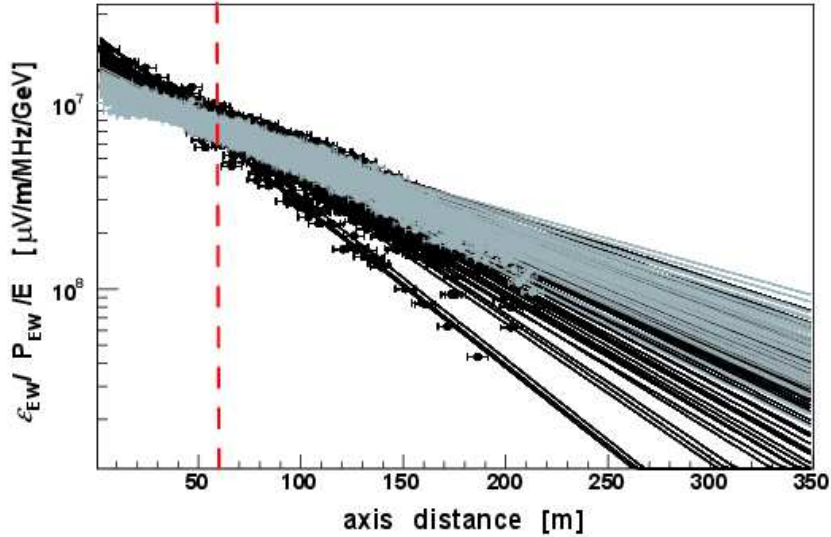}
  \caption{Normalized REAS3 lateral distribution function for the events in the first $\Delta \theta$, simulated once as proton (black), once as iron (gray). The point are fitted with eq.\ref{expf}}
   \label{LDF}
\end{figure} 
The radio detection technique offers two different approaches for the investigation of the primary composition: by measuring the radio wavefront shape \cite{frank} and by looking at the LDF slope (slope-method).
The slope in the radio LDF depends also on the shower inclination. For inclined showers the radio source is further away compared to a vertical shower, and this implies a flatter LDF slope. To reduce the zenith angle dependence of the LDF slope, and better focus on the primary composition effects, five zenith angle bins are thus considered.\\ 
The LDF for the events in the first zenith angle bin, simulated as both proton- and iron- initiated showers, are shown in fig.\ref{LDF}. 
The electric field in the antennas is normalized by the primary energy and by the arrival direction (P$_{\mathrm{EW}}$) of the air shower. 
P$_{\mathrm{EW}}$ is the component along the east-west direction of the normalized Lorentz force 
vector \overrightarrow{P} $/\mid$\overrightarrow{v}$\cdot$\overrightarrow{B}$\mid$ $\simeq$ sin($\alpha$), with \overrightarrow{v} the velocity of the primary particle and \overrightarrow{B} the geomagnetic field. This is known to be just a simple approximation, first because the geomagnetic effect is known not to be the only radio emission mechanism, second because the LOPES antennas partially detect also the vertical component of the total electric field. Requiring P$_{\mathrm{EW}}$ $>$ 0.2 is thus necessary in order to reject the few events for which the geomagnetic contribution to the radio emission is small.\\
The so-called \textit{flat} region (d$_{flat}$), mainly the distance from the shower axis at which the detected electric field in the antenna ($\mathrm{\epsilon_{flat}}$) does not depend on the primary particle type, is identified. At d$_{flat}$ the $\mathrm{\epsilon_{flat}}$ normalized by the primary energy and direction of the shower is expected to be almost the same for all the events and the LDFs to intersect in one point. In this analysis the \textit{flat} region is identified where the relative spread of all the LDF fits is at the minimum and this occurs at around 60$\pm$5 m for the first zenith bin. The \textit{flat} regions for all zenith angle ranges are listed in Table \ref{table_flat}, suggesting that the slope-method can be easily applied to the LOPES data, since they are predicted inside the LOPES array. Anyway it should be stressed that the value for the d$_{flat}$ depends both on the observing frequency and the altitude of the experiment \cite{REAS2c}, so it is a LOPES-specific value. \\
From the simulations one expects the charge excess contribution to the radio emission to strongly depend on the azimuthal observer position with respect to the shower core \cite{REAS3,MGMRc}. Since REAS3 includes this contribution and, moreover, the simple one-dimensional-exponential fitting function (eq.\ref{expf}) is used, one could expect a high value even for the minimum dispersion of the LDFs fit. 
This is actually quite small, around 6-8$\%$ in the complete zenith range (Table \ref{table_flat}), and it is the uncertainty on the electric field mostly due to the shower-to-shower fluctuations.\\ 
On the contrary the $steep$ region is another distance from the shower axis, where the electric field ($\mathrm{\epsilon_{steep}}$) brings also the information about the primary mass. For the LOPES setup the $\mathrm{\epsilon_{steep}}$ is chosen to be at 170 m far from the $\mathrm{d_{flat}}$. The $\mathrm{\epsilon_{flat}}$/$\mathrm{\epsilon_{steep}}$ ratio of each event can be used as indicator of the radio LDF slope. 

\begin{table}[t]
\begin{center}
\begin{tabular}{l|c|c|c}

$\Delta \theta$ & d$_{flat}$ [m] &  R$_{flat}$ [m] &  RMS$_{min} \%$  \\ 
\hline
0.$^{\circ}$-19.4$^{\circ}$     & 60 & $\leq$ 64  & 6.0\\
19.4$^{\circ}$-26.8$^{\circ}$   & 70 & $\leq$ 67  & 6.2\\
26.8$^{\circ}$-32$^{\circ}$     & 70 & $\leq$ 83  & 6.4\\
32$^{\circ}$-36.2$^{\circ}$     & 90 & $\leq$ 111  & 7.6\\
36.2$^{\circ}$-40$^{\circ}$     & 90 & $\leq$ 117 & 8.0\\
\end{tabular}
\caption{Distance of the $flat$ region and RMS of the electric field values at d$_{flat}$. R is the distance from the shower core, in the ground based coordinate system, while d is the equivalent value in the shower plain coordinate system.}\label{table_flat}
\end{center}
\end{table}

\begin{figure}[h!]
  \centering
  \includegraphics[width=5.in,height=3.in]{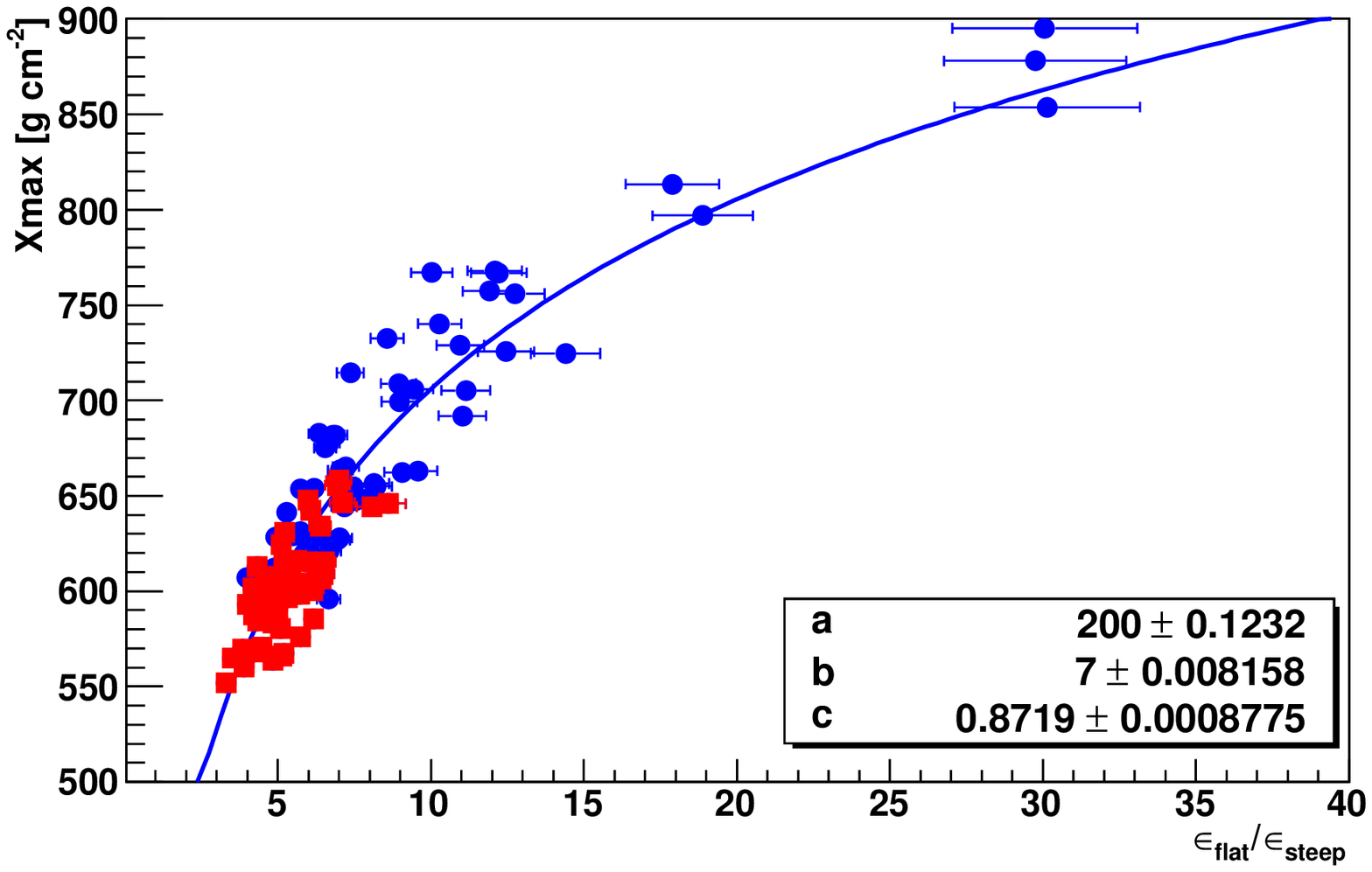}
  \caption{Correlation between the true CORSIKA X$_{\mathrm{max}}$ and the radio simulated LDF slope, for proton (blue) and iron (red) simulated primaries.}
  \label{xmax}
  \hspace{4cm}
  \centering
  \includegraphics[width=5.in,height=3.in]{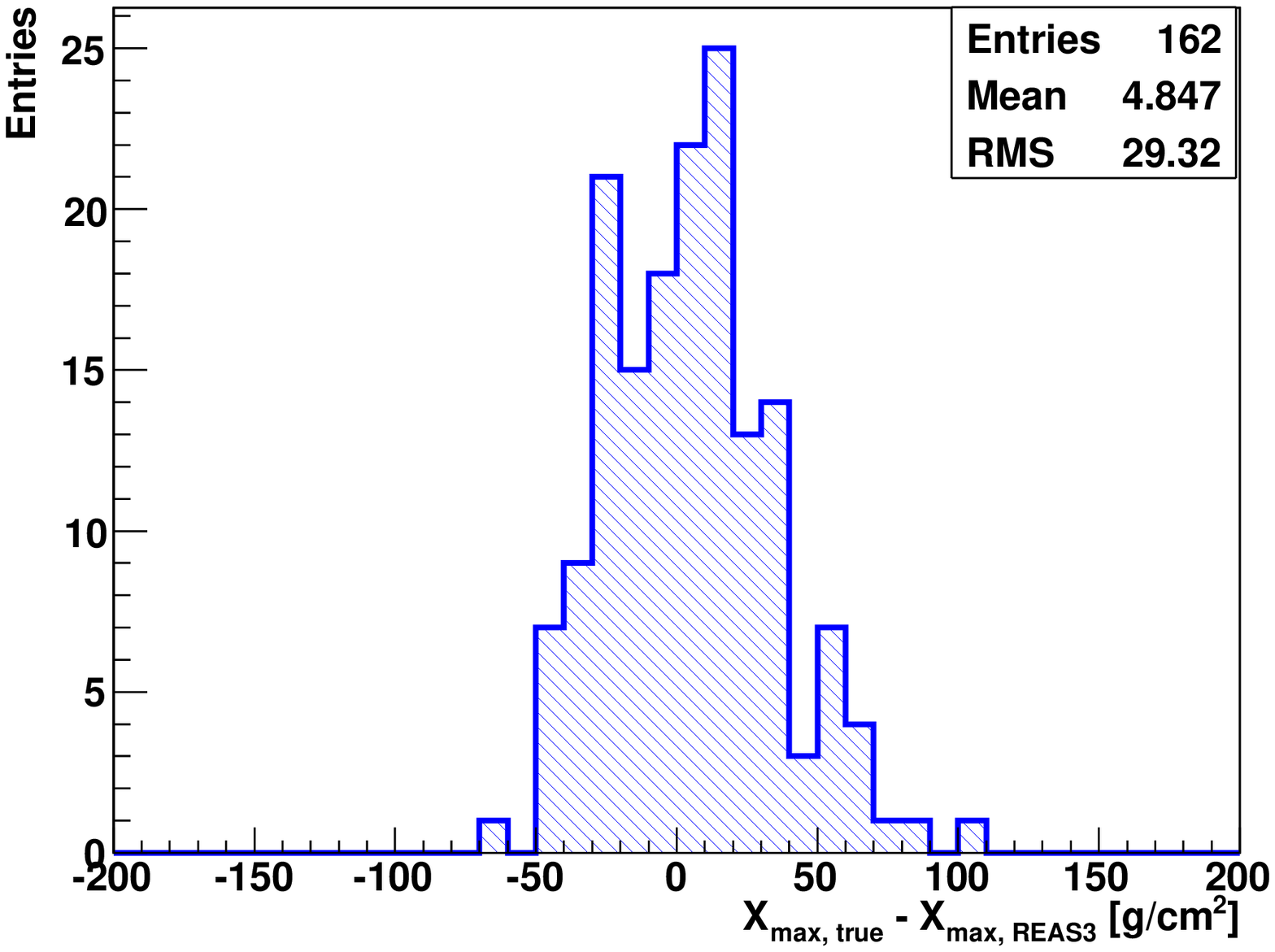}
   \caption{Dispersion of the points (X$_{\mathrm{max,true}}$) around the fit (X$_{\mathrm{max,REAS3}}$) of fig.\ref{xmax}}
  \label{xmaxdisp}
\end{figure}
\begin{figure}[h!]
 \vspace{2mm}
 \centering
 \includegraphics[width=5.in]{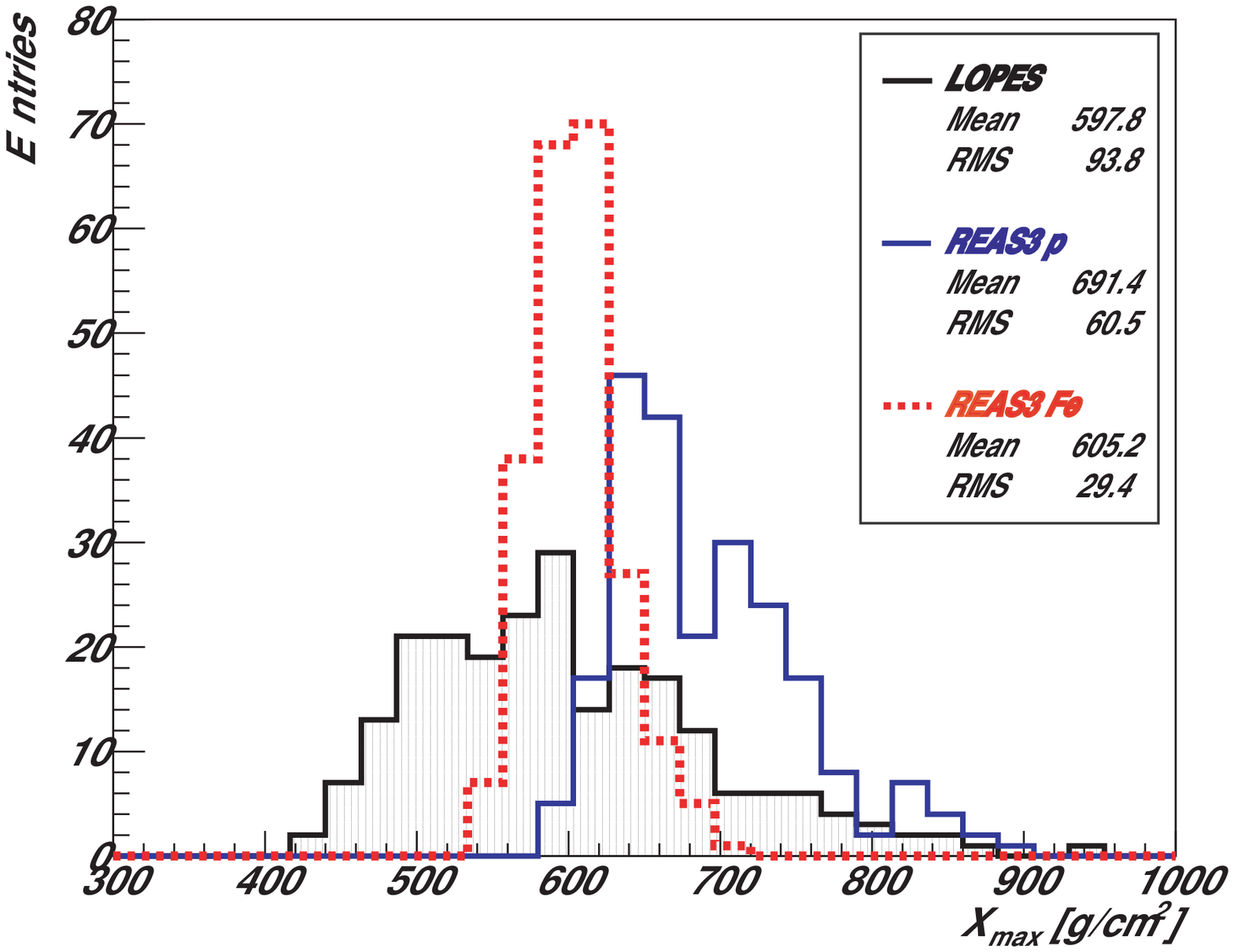}
 \caption{LOPES reconstructed X$_{\mathrm{max}}$ using the slope-method procedure.}
 \label{xmaxHisto}
\end{figure}
\section{Xmax reconstruction for simulations and LOPES data}
 
To verify the correlation between the radio LDF slope and the primary composition, the $\mathrm{\epsilon_{flat}}$/$\mathrm{\epsilon_{steep}}$ ratio of each event is compared with the corresponding X$_{\mathrm{max}}$ from the CORSIKA simulation. This will be here referred to as the $true$ value for the air shower X$_{\mathrm{max}}$.
The error on $\mathrm{\epsilon_{flat}}$/$\mathrm{\epsilon_{steep}}$ comes from the Gaussian error propagation of the uncertainties on the LDF-fit parameter R$_{0}$ and on the d$_{flat}$. For the events in the first zenith angle bin this correlation is shown in fig.\ref{xmax} and the points are fitted with function
\begin{equation}
X_{max} = a \left[\mathrm{ln} \left(b \frac{\epsilon_{\mathrm{flat}}}{\epsilon_{\mathrm{steep}}}\right)\right]^{c}
\label{xm_eq}
\end{equation}
as in \cite{REAS2c}; this nice behavior is similarly seen for all the other zenith bins.
The dispersion of the points around the fit of fig.\ref{xmax} is calculated and the RMS spread of its distribution is circa 29 g/cm$^{2}$ (fig.\ref{xmaxdisp}), which can be referred to as the uncertainty of the slope-method in reconstructing X$_{\mathrm{max}}$. In the whole zenith angle range, a maximum uncertainty of 40 g/cm$^{2}$ is found for the most inclined showers. \\
One has to point out that the proton simulated events (points), which have larger values for  X$_{\mathrm{max,true}}$ and for $\mathrm{\epsilon_{flat}}$/$\mathrm{\epsilon_{steep}}$, have also larger values for the $\mathrm{\epsilon_{flat}}$/$\mathrm{\epsilon_{steep}}$ errors. This implies that the larger weight to the fit comes from the iron-initiated showers (squares), which fluctuate much less.\\ 
The X$_{\mathrm{max, REAS3}}$ reconstructed with the slope-method for the proton and iron REAS3-simulated showers, i.e. the values of the fit in fig.\ref{xmax}, are shown in fig.\ref{xmaxHisto}.\\
Eq. \ref{xm_eq} can be directly applied to reconstruct the experimental X$_{\mathrm{max, LOPES}}$ for the LOPES data. The parameters $a$, $b$, and $c$, in the formula applied, are the same obtained from the simulations, in each $\Delta \theta$. The $\mathrm{\epsilon_{flat}}$/$\mathrm{\epsilon_{steep}}$ information, on the other side, comes  directly from the LOPES data LDF fit.\\
The X$_{\mathrm{max, LOPES}}$, reconstructed for all the data selection, are presented in fig.\ref{xmaxHisto}, with a mean and a standard deviation value of $\sim$ 600 $\pm$ 90 g/cm$^{2}$. In the real case of a mixed composition selection, the statistical uncertainty of the LOPES X$_{\mathrm{max, LOPES}}$ is comparable with the REAS3 prediction. Moreover, the LOPES reconstructed   
X$_{\mathrm{max}}$ values are quite compatible with the expectations from the cosmic ray nuclei.\\
However, the LOPES reconstructed values are slightly shifted to X$_{\mathrm{max}}$ smaller than the REAS3 iron-like expectations and this systematic is surely influenced by the hadronic interaction model used for this set of simulations (QGSJetII).

\newpage
\section{Conclusion}
An investigation on the expected correlation between the radio LDF slope and the mass of the primary cosmic ray is now possible due to the more complete REAS3 code and to the general agreement with the LOPES data.\\ 
The slope-method has been tested on REAS3 simulations of LOPES selected events, revealing the possibility to directly apply the analysis to the experimental data.\\
X$_{\mathrm{max,LOPES}}$ values are reconstructed with the LOPES LDF information and are quite compatible with the expectations for the cosmic ray nuclei. 
A general agreement is also seen with the REAS3 reconstructed X$_{\mathrm{max,REAS3}}$, even though there is a shift to values smaller than the iron initiated showers prediction, which, anyway, is influenced by the hadronic interaction model specifically used in this set of simulations.\\  
One must also take into account that the values for the $flat$ region, important for such investigation and found for this analysis, are restricted to the LOPES simulated setup, so they are not directly applicable to other radio experiments.\\
The possibility to reach a radio mass sensitivity also through the detected lateral distribution function slope was shown.\\
In the future, applying the slope-method to radio detector experiments located in a lower noise background could highly improve the resolution on the reconstructed X$_{\mathrm{max}}$, so on the sensitivity on the primary mass. Especially in the case of AERA (Auger Engineering Radio Array), located at the Pierre Auger Observatory, which could cross-check the reconstructed X$_{\mathrm{max}}$ with the experimental values obtained by the Fluorescence Detector (FD).


\vspace{\baselineskip}
\

\clearpage

\end{document}